      [1999/12/01 v1.4c Il Nuovo Cimento]
\documentclass{cimento}

\usepackage{slashed,amsmath,amssymb,wrapfig} 
\usepackage{epsfig}
\newcommand{\beq} {\begin{equation}}
\newcommand{\eeq} {\end{equation}}
\newcommand{\beqa} {\begin{eqnarray}}
\newcommand{\eeqa} {\end{eqnarray}}

\newcommand{\as}{{\alpha_s}}

\newcommand{\chat}{\hat{\bs{\ell}}}
\newcommand{\la}{\Lambda}

\newcommand{\ieps}{i\varepsilon}

\newcommand{\order}[1]{${\cal O}\left(#1 \right)$}

\newcommand{\halft}{{\textstyle \frac{1}{2}}}

\newcommand{\ket}[1]{\left\vert{#1}\right\rangle}
\newcommand{\bra}[1]{\langle{#1}\vert}

\newcommand{\bs}[1]{\boldsymbol{#1}}

\newcommand{\xv}{{\bs{x}}} 

\newcommand{\pv}{{\bs{p}}}

\newcommand{\nv}{\bs{\nabla}}

\title{Are hadrons simpler than they seem?}
\author{Paul Hoyer\from{ins:x}% \ETC,
\thanks{Presented at the {\it 3rd Workshop on the QCD Structure of the Nucleon} (QCD-N'12), 22-26 October 2012 in Bilbao, Spain. Based on work done with D. D. Dietrich and M. J\"arvinen \cite{Dietrich:2012iy}.}}
\instlist{\inst{ins:x} Department of Physics and Helsinki Institute of Physics\\ POB 64, FIN-00014 University of Helsinki, Finland}
\begin{document}

\maketitle

\begin{abstract}
I briefly review a systematic approximation scheme of QCD in which the quark model picture of hadrons emerges at lowest order. A linear $A^0$ potential arises if Gauss' law is solved with a non-vanishing boundary condition at spatial infinity. Similarly to the Dirac case one can describe relativistic states including any number of particle pairs (sea quarks) using valence wave functions, whose norms give {\em inclusive} probability densities. Provided $\as(Q^2)$ freezes in the infrared, perturbative corrections to the $S$-matrix can be calculated in the usual way, but with states bound by the linear \order{\as^0} potential instead of plane waves in the $in$ and $out$ states.

\vspace{5mm}


\vspace{-5mm}

\end{abstract}

\section{Questions and Answers}

Hadrons have two seemingly incompatible features:
\begin{itemize}

\item Their (light) quark constituents are highly relativistic, consequently hadron wave functions have abundant contributions from sea quarks and gluons. This is seen in the parton distributions measured in hard processes, and is required by the underlying theory of Quantum Chromodynamics (QCD).

\item The hadron spectrum reflects just the valence quark ($q\bar q, qqq$) degrees of freedom.  Sea quarks and gluons do not manifest themselves as additional states (multi-quark states, hybrids, glueballs). The spectra of heavy quarkonia are remarkably similar to the atomic spectra of perturbative QED.

\end{itemize}

This raises the following questions:

\begin{enumerate}

\item {\it What approximation allows relativistic field theory bound states, with prominent multi-particle Fock components, to be described by valence quark wave functions?}\\
The bound states of electrons in a strong Coulomb field have Fock components with multiple $e^+e^-$ pairs, yet are described by Dirac wave functions that have only a single electron degree of freedom. I discuss why this implies that the norm of the wave function is an inclusive probability density. I also recall a remarkable feature of the Dirac wave functions which has been largely forgotten since the early 1930's.  

\item {\it How does the quark model potential $V(r)=c\,r-C_F\alpha_s/r$ arise in QCD?}\\
The linear term stems from a non-vanishing boundary condition in solving Gauss' law for $A^0$. The single gluon exchange term is an \order{\as} correction to the linear term. The perturbative expansion is meaningful provided the strong coupling freezes at low momentum transfers, with several estimates indicating $\alpha_s(0) \simeq 0.5$. 

\end{enumerate}

\section{Relevant aspects of the Dirac wave function}

The Born level Green function $G_B$ of an electron in an external $A^0(\xv)$ field generated by a static charge $eZ$ is obtained by summing all Coulomb exchanges between the electron and the charge. The residue $R_B$ of a bound state pole, $G_B(p^0,\pv) = R_B(E,\pv)/(p^0-E)$, satisfies the Dirac equation with potential $eA^0(\xv)$. In this Born approximation all \order{e^2} vertex corrections as well as loops created by the external field are ignored. Since the field is static it does not change the $p^0$ component of the electron momentum. If $p^0 > -m$ the $\ieps$ regularization is irrelevant at the negative energy poles of the electron propagators, since $p^0+\sqrt{\pv^2+m^2}>0$.

In order to find the wave function of a bound state at an instant of time $t$ we need to time-order the interaction vertices. Electrons of negative energy move backward in time, creating so called `$Z$-diagrams' where one (or several) additional $e^+e^-$ pairs are propagating along with the electron. An equal-time, relativistic bound state wave function necessarily contains any number of such pairs. As the strength of the potential increases the bound state becomes more relativistic and the pairs become more prominent.

The usual Dirac wave function $\psi(t,\xv)=\exp(-iEt)\psi(\xv)$ depends on the position $\xv$ of a single electron, which may have either positive or negative energy. This wave function is obtained when retarded, rather than Feynman electron propagators are used. As we noted above, the Green function $G_B(p^0,\pv)$ is independent of the $\ieps$ prescription at the negative energy poles of the electron propagators. In particular, the bound state energies $E$ are the same whether retarded or Feynman electron propagators are used. However, the equal-time wave function is obtained after Fourier transforming $p^0 \to t$, and is sensitive to the $\ieps$ prescription. In retarded propagation also negative energy electrons move forward in time, there are no $Z$-diagrams and the amplitude of the single electron present at time $t$ is given by $\psi(t,\xv)$.

The fact that the Dirac wave function $\psi(t,\xv)$ is obtained using retarded (rather than Feynman) boundary conditions means, in analogy to cross sections \cite{Baltz:2001dp}, that $|\psi(t,\xv)|^2$ should be interpreted as an {\em inclusive} probability density. As shown by Weinberg \cite{406190}, the norm of the Dirac wave function should  be normalized to unity {\em provided the normalizing integral converges}. Already in the early 1930's it was realized \cite{plesset} that the normalization integral diverges for most potentials -- the $1/r$ potential of $D=3+1$ dimensions being an exception. {\it E.g.,} for the linear potential $V(x) = \halft e^2Z|x|$ of QED in $D=1+1$ the norm $|\psi(t,x)|^2$, and hence the inclusive particle density, is constant at large distances $x$. This is consistent with the idea of string breaking, particle pairs being created over an interval of $x$ where the potential increases by $2m$. It is also in accord with the behavior of the Dirac wave function for nearly non-relativistic dynamics, $e^2Z \ll m^2$ \cite{Dietrich:2012iy}. The normalization integral being divergent furthermore implies that the Dirac spectrum is {\em continuous} rather than discrete \cite{plesset}. Curiously, these properties of the Dirac solutions have been scarcely discussed since they were first noticed long ago.

\section{The linear potential}

The linear confining potential of the quark model is necessary for describing the hadron spectrum. This introduces a dimensionful constant $\sim \la_{QCD}$, which is not present in the QCD Lagrangian. Since the Schr\"odinger equation is a Born level approximation the constant cannot originate from the renormalization of loop integrals. The only possibility consistent with the underlying theory is then to introduce it via a boundary condition. 

In gauge theories the Coulomb potential $A^0$ is instantaneous, since its time derivative does not appear in the Lagrangian. At each instant of time the charge distribution determines $A^0$ through the field equations of motion. {\it E.g.,} in the Hydrogen atom of QED, for each position $\xv_1$ of the electron and $\xv_2$ of the proton we find $A^0(\xv;\xv_1,\xv_2)$ by solving Gauss' law, $-\nv^2_{\xv}A^0=e[\delta^3(\xv-\xv_1)-\delta^3(\xv-\xv_2)]$. The standard solution gives rise to the Coulomb potential $V(\xv_1-\xv_2)=-\alpha/|\xv_1-\xv_2|$ when both the field energy $\halft\int d^3\xv (\nv A^0)^2$ and the interaction energies $eA^0(\xv_1)-eA^0(\xv_2)$ are taken into account (the infinite self-energies are independent of $\xv_1,\xv_2$ and thus irrelevant).

The $-\alpha/r$ Coulomb potential of the Hydrogen atom results when Gauss' law is solved with the boundary condition that $A^0(\xv;\xv_1,\xv_2) \to 0$ as $|\xv| \to\infty$. If we instead require that the field strength approaches a non-vanishing constant, $(\nv A^0)^2 \to \Lambda^4$ as $|\xv| \to\infty$, then we have to add a homogeneous solution to Gauss' law, $A^0_\Lambda = \Lambda^2\,\chat\cdot \xv$, where $\chat(\xv_1,\xv_2)$ is a unit vector. The square of $\nv A^0_\Lambda$ contributes a term $\propto\Lambda^4 V$ to the field energy ($V$ is the volume of space). This term is irrelevant provided $\la$ is a universal constant, independent of $\xv_1,\xv_2$. The field energy arising from the interference of $A^0_\Lambda$ with the \order{e} potential is finite provided the state is neutral, and is $\propto e\la^2 \chat\cdot(\xv_1-\xv_2)$. Choosing $\chat\parallel \xv_1-\xv_2$ makes the action stationary under variations of the unit vector $\chat$, preserves rotational invariance and gives rise to the linear potential $V_\la \propto e\la^2|\xv_1-\xv_2|$.

A linear potential arises also in QCD through an analogous boundary condition \cite{Hoyer:2009ep}. This does not explain why we should choose $\la = 0$ in QED and $\la \sim \la_{QCD} \neq 0$ in QCD to describe observed phenomena. However, it shows that a linear potential is {\em compatible} with the field equations of motion. $\la$ is a hidden parameter, not present in the Lagrangian, whose value must be determined by experiment. As indicated above, the solution requires $\la$ to be a universal constant and the state to be charge neutral. In QCD bound state solutions were found \cite{Hoyer:2009ep} for color neutral $q\bar q$ mesons and $qqq$ baryons. No analogous solutions exist for states with a higher number of valence quarks.

\section{Born level bound states}

The linear $A^0$ potential ensures confinement and relativistic binding. Since the potential is instantaneous hadrons may be described using ``inclusive'' valence quark wave functions analogous to those of Dirac states. Perturbative corrections can then be added systematically. In the $S$-matrix expression 
%\beq\label{smatrix}
$S_{fi}={}_{\rm out}\bra{f}T\big[\exp(-i \int d^4x\,\mathcal{H}_I)\big]\ket{i}_{\rm in}$
%\eeq
the $in$ and $out$ states should consist of neutral states bound by the linear potential, instead of the free plane wave states of standard perturbation theory. In effect, one expands around lowest order states which already have the basic features of hadrons such as color confinement. As in the Taylor expansion of ordinary functions, the full perturbative series formally gives the exact result independently of the starting configuration.

For the perturbative expansion to be meaningful the strong coupling should freeze at a perturbative value in the infrared. Phenomenological and theoretical studies \cite{Brodsky:2002nb} find $\as(0) \simeq 0.5$, which justifies the use of perturbation theory. 

The Poincar\'e invariance of the $S$-matrix requires that the $in$ and $out$ states transform covariantly. Equal-time, interacting states transform dynamically under boosts, since the concept of equal time is frame dependent. The Poincar\'e invariance of the action, and the Lorentz invariant formulation $F_{\mu\nu}F^{\mu\nu}=-2\la^4$ of the boundary condition, suggests that the bound states should transform correctly under boosts. A detailed analysis in $D=1+1$ dimensions showed that the boost operator indeed transforms the states as required. From this follows, in particular, that the bound state energy $E$ has the correct dependence on the CM momentum, $E=\sqrt{M^2+P^2}$. The same dependence was found previously in $D=3+1$ \cite{Hoyer:1986ei}. The covariance holds only for an exactly linear potential. Under boosts the $D=1+1$ wave function contracts $\propto 1/[E-V(x)]$. In the non-relativistic limit $V \ll E$ this reduces to the usual Lorentz $\gamma$-factor, but in general the wave function contracts at a rate which depends on the separation $x$ of the constituents. In effect, the relevant quantity is thus found to be the kinetic energy $p^0-eA^0$ rather than the canonical energy $p^0$ familiar from classical physics.

For a more complete list of references and a discussion of the wave functions, form factors and parton distributions of the bound states in $D=1+1$ see \cite{Dietrich:2012iy}.

\acknowledgments
I thank the QCD-N'12 workshop organizers for the possibility to present this work, and the Magnus Ehrnrooth Foundation for a travel grant. I am grateful to the Institute for Nuclear Theory at the University of Washington for its hospitality, and to the Department of Energy for partial support during the completion of this note.

\end{document}